\documentclass[10pt]{article}
\usepackage{fullpage,amssymb,amsmath}
\usepackage[dvips]{epsfig}

\def\##1{{\bf #1}}
\def\=#1{\underline{\underline{#1}}}

\def\+#1{\underline{\bf #1}}
\def\*#1{\underline{\underline{\bf #1}}}

\def\r#1{(\ref{#1})}
\def\l#1{\label{#1}}
\def\c#1{\cite{#1}}

\def\le{\left(}
\def\ri{\right)}
\def\les{\left[}
\def\ris{\right]}
\def\lec{\left\{}
\def\ric{\right\}}

\def\.{\cdot}

\def\eps{\epsilon}

\begin{document}

\begin{center}

{\bf {\LARGE On the sensitivity of generic porous optical sensors}}

\vspace{10mm} \large

 Tom G. Mackay\footnote{E--mail: T.Mackay@ed.ac.uk}\\
{\em School of Mathematics and
   Maxwell Institute for Mathematical Sciences\\
University of Edinburgh, Edinburgh EH9 3JZ, UK}\\
and\\
 {\em NanoMM~---~Nanoengineered Metamaterials Group\\ Department of Engineering Science and Mechanics\\
Pennsylvania State University, University Park, PA 16802--6812,
USA}\\

\end{center}

\begin{abstract}

A porous  material was considered as a platform for  optical
sensing. It was envisaged that the porous material was infiltrated
by a fluid which contains an agent to be sensed. Changes in the
optical properties of the infiltrated porous material provide the
basis for detection of the agent to be sensed. Using a
homogenization approach based on the Bruggeman formalism, wherein
the infiltrated porous material was regarded as a homogenized
composite material, the sensitivity of such a sensor was
investigated. For the case of an isotropic dielectric porous
material of relative permittivity $\eps^a$ and an isotropic
dielectric fluid of relative permittivity $\eps^b$, it was found
that the sensitivity was maximized when there was a large contrast
between $\eps^a$ and $\eps^b$; the maximum sensitivity was achieved
at mid-range values of porosity. Especially high  sensitivities may
be achieved for $\eps^b$ close to unity when $\eps^a \gg 1$, for
example. Furthermore, higher sensitivities may be achieved by
incorporating pores which have elongated spheroidal shapes.

\end{abstract}

%\ocis{280.4788, 260.2065, 160.1245}

%\maketitle %% null function with osajnl.sty

\section{Introduction}

A simple generic optical sensor may be envisaged as porous  material
(labelled $a$, say), which is infiltrated by a fluid (labelled  $b$,
say). An agent to be sensed is contained within the fluid. It
assumed that the fluid and the agent to be sensed have quite
different optical properties. Thus, the concentration
 of the agent within the fluid may be gauged by the optical
 properties of the infiltrated porous material \c{Stefano,Pinet,Sipe_OE}.
 The optical properties
 used to detect the presence of the agent  may be the reflectances or transmittances
 of the infiltrated porous material. Alternatively, if one surface of
 the porous material
 were coated with a thin metallic film, measurements could be based
 on the excitation of surface-plasmon-polariton (SPP) waves at the
 interface of the porous material and metal film \c{Homola2003,AZL2,Scarano}.

For example, sculptured thin films (STFs) represent  rather
promising porous materials for such optical sensors
\c{LMBR,Lgas,L_Optik}. These constitute parallel arrays of nanowires
which are grown on substrates by physical vapour deposition
\c{STF_Book,Messier}. By controlled manipulation of the substrate
during the deposition process, a range of nanowire shapes can be
achieved. Thereby,
 the multiscale porosity of such  STFs
 can be tailored to order, to a considerable degree. Additionally, since
  STFs can fabricated from a wide range of organic and inorganic
  materials, a wide range of
  optical properties for the porous material can be delivered \c{Polo,Horn}.
 Chiral STFs are
 especially interesting for optical sensing applications,
 as these support the circular Bragg phenomenon, courtesy of
 the helical nature of their nanowires \c{STF_Book}; furthermore, they also support more than
 one mode of SPP wave \c{Polo_PRSA,DPL,PML_JOSAB} which may be usefully exploited for sensing \c{ML_IEEE_SJ}.

In the design of  such an optical sensor, what values should one
choose for the optical properties of the porous material and
infiltrating fluid,
 in order to maximize
sensitivity? What value should one choose for the porosity, and what
shape should one choose for the pores, in order to maximize
sensitivity?
 These are the questions that we address here. We do so by
considering the simplest scenario wherein the porous material and
infiltrating fluid are both made from lossless, homogeneous,
isotropic dielectric materials, characterized by relative
permittivities $\eps^a$ and $\eps^b$, respectively.\footnote{The
entire analysis presented herein may also be applied to lossless,
homogeneous, isotropic magnetic materials, by replacing relative
permittivities by the corresponding relative permeabilities
throughout.} The infiltrated porous material is regarded as a
homogeneous composite material (HCM), which is a reasonable
approximation provided that the linear dimensions of the pores  are
much smaller than the wavelength(s) involved. Thus, in the case of
optical sensors operating at visible wavelengths, we have in mind
pore linear dimensions $\lessapprox 38$ nm for the smallest values
of $\eps^{a,b}$ considered and $\lessapprox 10$ nm for the largest
values of $\eps^{a,b}$ considered.
 The
infiltrated porous material may be either isotropic or anisotropic
depending upon the shape of the pores. We use the well-established
Bruggeman homogenization formalism to estimate the relative
permittivity dyadic of the infiltrated porous material, namely
$\=\eps^{Br}$ \c{Ward,M_JNP}. The Bruggeman formalism has recently
been implemented to study the prospects of infiltrated STFs as
optical sensors \c{ML_inverse_homog}, based on  both changes in
reflectance/transmittance  \c{ML_IEEEPJ} and SPP wave excitation
\c{ML_IEEE_SJ,ML_PNFA}.

Regardless of whether the  sensor is based on changes in
reflectance/transmittance or  the excitation of SPP waves, the
sensitivity of the sensor depends crucially on how much the optical
properties of the infiltrated porous material change in response to
changes in the optical properties of the infiltrating fluid. Thus,
the derivative $d \=\eps^{Br} / d \eps^b$ is a key indicator of
sensitivity. In the following we  explore how this derivative varies
as a function of the porosity, the pore shape and the relative
permittivities of the infiltrating fluid and the porous material.

\section{Homogenization theory}

\begin{figure}[!htb]
\begin{center}
\begin{tabular}{c}
\includegraphics[width=12.0cm]{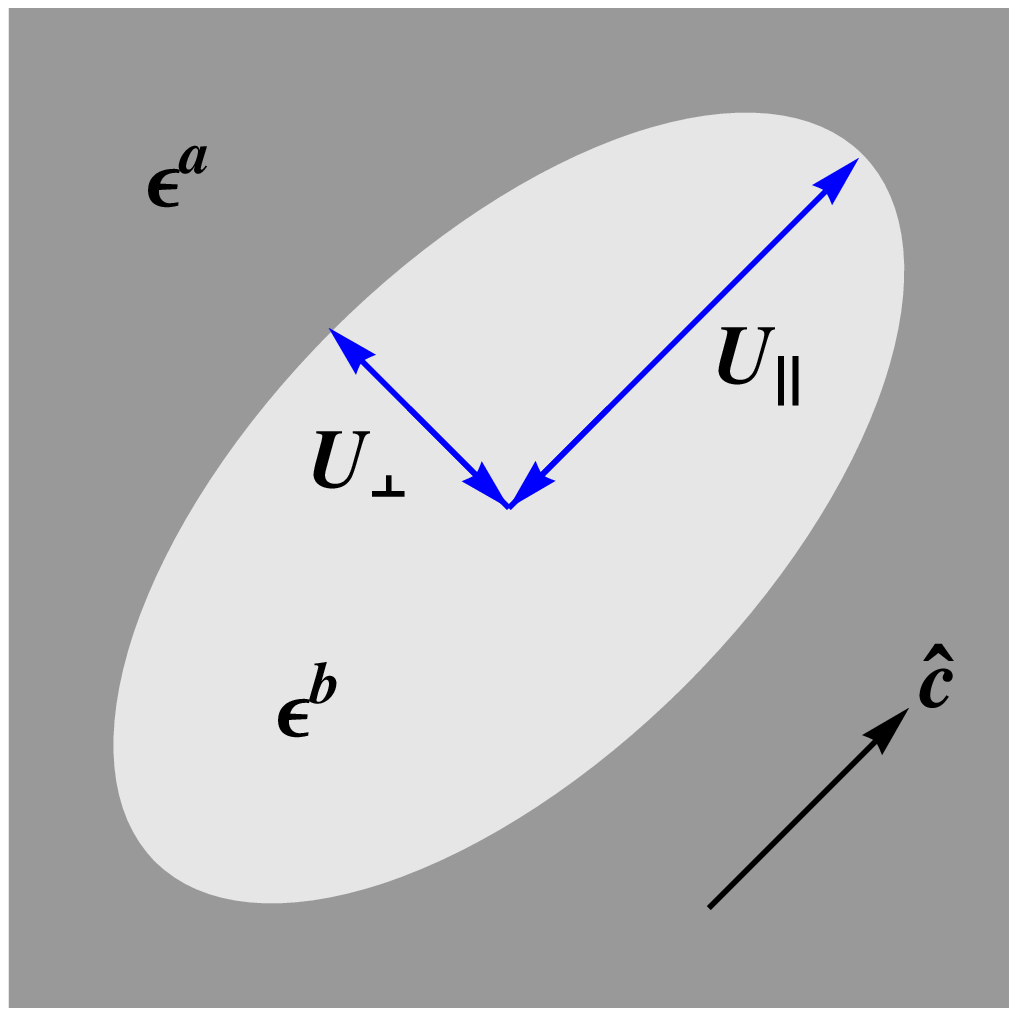}
\end{tabular}
\end{center}
 \caption{Schematic illustration of a spheroidal pore filled with a fluid of relative
 permittivity $\eps^b$, embedded in a material of relative
 permittivity $\eps^a$. The spheroid's semi-major and semi-minor axes
  have lengths $U_\parallel$ and $U_\perp$, respectively, with the
semi-major axis being aligned with the direction of $\hat{\#c}$,
  per eqn.~\r{Ushape}.
 } \label{fig1}
\end{figure}

Within our homogenization framework, the pores  are all assumed to
have the same   shape, which is spheroidal in general. These
spheroidal pores are randomly distributed but identically oriented.
The surface
 of each spheroid relative to its centroid is prescribed by
the vector \c{M_JNP}
\begin{equation}
\#r_{\,s} (\theta, \phi) = \eta \, \=U \. \hat{\#r} (\theta, \phi),
\end{equation}
with $ \hat{\#r} $ being the radial unit vector originating from the
spheroid's centroid, specified by the spherical polar coordinates
$\theta$ and $\phi$. The linear dimensions of the spheroid, as
determined by the parameter $\eta$, are  assumed to be small
relative to the electromagnetic wavelength(s).
 The spheroidal shape is captured by the dyadic
\begin{equation} \l{Ushape}
 \=U =  U_\perp \=I + \le  U_\parallel - U_\perp \ri \, \hat{\#c} \, \hat{\#c}\,,
\end{equation}
where $\=I$ is the identity 3$\times$3 dyadic and the unit vector
$\hat{\#c}$ is parallel to the spheroid's axis of rotational
symmetry. The  linear dimension parallel to $\hat{\#c}$, relative to
the equatorial radius of the spheroid, is provided by the shape
parameter $\rho = U_\parallel / U_\perp$. A schematic illustration
of such a spheroidal pore is provided in Fig.~\ref{fig1}.

The form of the relative permittivity dyadic $\=\eps^{Br} $, as
estimated using the Bruggeman homogenization formalism,  mirrors
that of the shape dyadic $\=U$. That is, it has the uniaxial form
\begin{equation}
\=\eps^{Br} =  \eps^{Br}_\perp \=I + \le \eps^{Br}_\parallel -
\eps^{Br}_\perp \ri\, \hat{\#c} \, \hat{\#c}. \l{eps_Br}
\end{equation}
It  emerges  as the solution of the dyadic Bruggeman equation
\c{WLM}
\begin{equation}
f_a \, \=\alpha^{a} + f_b \, \=\alpha^{b}  = \=0\,, \l{Br}
\end{equation}
where $\=0$ is the null 3$\times$3 dyadic. The scalars $f_a$ and
$f_b = 1 - f_a$ denote the respective volume fractions of the porous
material and infiltrating fluid. Thus, $f_b$ represents the porosity
of the optical sensor. The dyadics
\begin{equation}
\=\alpha^{\ell } = \le \eps^\ell \=I - \=\eps^{Br} \ri \.\les \, \=I
+
 \=D \. \le \eps^\ell \=I - \=\eps^{Br} \ri \,\ris^{-1}, \qquad
(\ell = a,b), \l{polar}
\end{equation}
are the polarizability density dyadics of the spheroids in the HCM,
while the depolarization dyadic $\=D$ in eqn.~\r{polar} is given by
the double integral \c{M97,MW97}
\begin{equation}
\=D = \frac{1}{ 4 \pi} \, \int^{2 \pi}_0 \, d \phi \, \int^\pi_0 \,
d \theta \, \sin \theta \, \le \frac{1}{
\hat{\#r}\.\=U^{-1}\.\=\eps^{Br}\.\=U^{-1}\.\hat{\#r}} \ri
\=U^{-1}\.\hat{\#r} \, \hat{\#r} \. \=U^{-1}\,. \l{depol}
\end{equation}
It may be expressed in the uniaxial form
\begin{equation}
\=D = D_\perp\, \=I + \le D_\parallel - D_\perp \ri \hat{\#c}
\,\hat{\#c}\,,
\end{equation}
with components
\begin{eqnarray}
D_\parallel &=& \frac{\gamma}{ \eps^{Br}_\parallel } \, \Gamma_\parallel ( \gamma ), \l{Dx}\\
D_\perp&=& \frac{1}{ \eps^{Br}_\perp} \, \Gamma_\perp (\gamma),
\l{D}
\end{eqnarray}
wherein the terms
\begin{eqnarray}
\Gamma_\parallel (\gamma) &=& \frac{1}{4 \pi}\, \int^{2 \pi}_0 \, d
\phi \, \int^\pi_0 \, d \theta \, \frac{\cos^2 \phi \sin^3
\theta}{\cos^2 \theta + \sin^2 \theta \le \gamma \cos^2 \phi +
\sin^2 \phi \ri}, \l{dx}\\
\Gamma_\perp(\gamma) &=&  \frac{1}{4 \pi}\, \int^{2 \pi}_0 \, d \phi
\, \int^\pi_0 \, d \theta \, \frac{\sin^2 \phi \sin^3 \theta}{\cos^2
\theta + \sin^2 \theta \le \gamma \cos^2 \phi + \sin^2 \phi \ri}
\l{d}
\end{eqnarray}
  are
functions of the scalar parameter
\begin{equation}
\gamma = \frac{U^2_\perp \eps^{Br}_\parallel}{U^2_\parallel
\eps^{Br}_\perp}.
\end{equation}
The double integrals on the right sides of eqns.~\r{dx} and \r{d}
may be evaluated as
\begin{eqnarray}
\Gamma_\parallel (\gamma )&=& \left\{
\begin{array}{lcr}
\displaystyle{
  \frac{\sinh^{-1} \sqrt{\frac{1
-\gamma}{\gamma} }}{\le 1 - \gamma  \ri^{\frac{3}{2}}} -
\frac{1}{1-\gamma }} && \hspace{14mm} \mbox{for} \;\; 0 < \gamma < 1
\\ & & \\
\displaystyle{ \frac{1}{\gamma - 1} - \frac{\sec^{-1} \sqrt{\gamma}
} {\le \gamma - 1 \ri^{\frac{3}{2}}}}& & \mbox{for} \;\; \gamma > 1
\end{array}
\right., \\
\Gamma_\perp( \gamma )&=& \left\{
\begin{array}{lcr}
\displaystyle{ \frac{1}{2} \le  \frac{1}{1-\gamma }-
  \frac{ \gamma \sinh^{-1} \sqrt{\frac{1
-\gamma}{\gamma} }}{\le 1 - \gamma  \ri^{\frac{3}{2}}} \ri } &&
\mbox{for} \;\; 0 < \gamma < 1
\\ & & \\
\displaystyle{\frac{1}{2} \le \frac{\gamma \sec^{-1} \sqrt{\gamma} }
{\le \gamma - 1 \ri^{\frac{3}{2}}} - \frac{1}{\gamma - 1}  \ri }& &
\mbox{for} \;\; \gamma > 1
\end{array}
\right..
\end{eqnarray}
Notice that the anomalous case $\gamma < 0$ which represents a
hyperbolic HCM \c{MLD2005} is excluded from our consideration.

The dyadic Bruggeman equation \r{Br} yields the two nonlinear scalar
equations
\begin{eqnarray}
&& \frac{\eps^a - \eps^{Br}_\parallel }{1 + D_\parallel \le \eps^a -
\eps^{Br}_\parallel \ri} f_a + \frac{\eps^b - \eps^{Br}_\parallel
}{1 + D_\parallel \le \eps^b - \eps^{Br}_\parallel \ri} f_b = 0 \,,
\l{Br_1}
\\
&& \frac{\eps^a - \eps^{Br}_\perp}{1 + D_\perp\le \eps^a -
\eps^{Br}_\perp\ri} f_a + \frac{\eps^b - \eps^{Br}_\perp}{1 +
D_\perp\le \eps^b - \eps^{Br}_\perp\ri} f_b = 0\,, \l{Br_2}
\end{eqnarray}
which are  coupled via $D_{\perp, \parallel}$. Using standard
numerical techniques, this pair can be solved  for
$\eps^{Br}_\parallel$ and $\eps^{Br}_\perp$.

Let us turn to the dyadic derivative which provides a measure of the
sensitivity of the porous optical sensor under consideration, namely
\begin{equation}
\frac{d \=\eps^{Br}}{d \eps^b}  =
 \frac{d \eps^{Br}_\perp}{d \eps^b}\, \=I + \le  \frac{d \eps^{Br}_\parallel}{d \eps^b} -
 \frac{d \eps^{Br}_\perp}{d \eps^b} \ri\, \hat{\#c} \, \hat{\#c}
\,. \l{deps_Br}
\end{equation}
Before proceeding further, we observe that the corresponding
derivatives of the depolarization dyadic components may be expressed
as
\begin{eqnarray}
\frac{d D_\parallel}{d \eps^b } &=&  \nu_{11} \frac{d
\eps^{Br}_\parallel}{d
\eps^b} + \nu_{12} \frac{d \eps^{Br}_\perp}{d \eps^b}\,,\\
\frac{d D_\perp}{d \eps^b}  &=& \nu_{21} \frac{d
\eps^{Br}_\parallel}{d \eps^b} + \nu_{22} \frac{d \eps^{Br}_\perp}{d
\eps^b}\,,
\end{eqnarray}
with the scalars
\begin{eqnarray}
\nu_{11} &=& \frac{U^2_\perp}{U^2_\parallel \eps^{Br}_\parallel
\eps^{Br}_\perp} \le \Gamma_\parallel + \gamma
\frac{d\Gamma_\parallel}{d \gamma} \ri - \frac{\gamma
\Gamma_\parallel}{\le
\eps^{Br}_\parallel \ri^2} \,,\\
\nu_{12} &=& - \frac{U^2_\perp }{U^2_\parallel \le \eps^{Br}_\perp\ri^2} \le \Gamma_\parallel + \gamma \frac{d\Gamma_\parallel}{d \gamma} \ri\,,\\
\nu_{21} &=& \le \frac{U^2_\perp}{U^2_\parallel \le
\eps^{Br}_\perp\ri^2} \ri \,
\frac{d\Gamma_\perp}{d \gamma}  \,,\\
\nu_{22} &=& - \le \frac{U^2_\perp
\eps^{Br}_\parallel}{U^2_\parallel \le \eps^{Br}_\perp\ri^3} \ri\,
\frac{d\Gamma_\perp}{d \gamma}  - \frac{ \Gamma_\perp}{\le
\eps^{Br}_\perp\ri^2} \,,
\end{eqnarray}
and derivatives
\begin{eqnarray}
\frac{d\Gamma_\parallel}{d \gamma} &=&
 \left\{
\begin{array}{lcr}
\displaystyle{ \frac{1}{2} \le
  \frac{3 \sinh^{-1} \sqrt{\frac{1
-\gamma}{\gamma} }}{\le 1 - \gamma  \ri^{\frac{5}{2}}} - \frac{1 + 2
\gamma}{ \le 1-\gamma \ri^2 \gamma } \ri } && \hspace{10mm}
\mbox{for} \;\; 0 < \gamma < 1
\\ & & \\
\displaystyle{\frac{1}{2} \le  - \frac{1 + 2 \gamma}{\le \gamma - 1
\ri^2 \gamma } + \frac{3 \sec^{-1} \sqrt{\gamma} } {\le \gamma - 1
\ri^{\frac{5}{2}}} \ri }& & \mbox{for} \;\; \gamma > 1
\end{array}
\right., \\
\frac{d \Gamma_\perp}{d \gamma} &=& \left\{
\begin{array}{lcr}
\displaystyle{ \frac{1}{4} \le   \frac{3}{ \le 1-\gamma \ri^2 } -
  \frac{\le 2 +  \gamma \ri \sinh^{-1} \sqrt{\frac{1
-\gamma}{\gamma} }}{\le 1 - \gamma  \ri^{\frac{5}{2}}} \ri } &&
\mbox{for} \;\; 0 < \gamma < 1
\\ & & \\
\displaystyle{\frac{1}{4} \le - \frac{ \le 2+ \gamma \ri \sec^{-1}
\sqrt{\gamma} } {\le \gamma - 1 \ri^{\frac{5}{2}}} + \frac{3}{\le
\gamma - 1 \ri^2}  \ri }& & \mbox{for} \;\; \gamma > 1
\end{array}
\right..
\end{eqnarray}
Next we exploit  the scalar Bruggeman equations \r{Br_1} and
\r{Br_2}. Their derivatives with respect to $\eps^b$ may be written
as
\begin{eqnarray}
&&  \beta_{11} \frac{d \eps^{Br}_\parallel}{d
\eps^b} + \beta_{12} \frac{d \eps^{Br}_\perp}{d \eps^b} + \beta_{13} = 0\,,\\
&& \beta_{21} \frac{d \eps^{Br}_\parallel}{d \eps^b} + \beta_{22}
\frac{d \eps^{Br}_\perp}{d \eps^b} + \beta_{23} = 0 \,,
\end{eqnarray}
with
\begin{eqnarray}
\beta_{11} &=& \nu_{11} \le \eps^a - \eps^{Br}_\parallel \ri \le
\eps^b - \eps^{Br}_\parallel \ri + D_\parallel \le 2
\eps^{Br}_\parallel - \eps^a - \eps^b \ri -1 \,, \\
\beta_{12} &=&
\nu_{12} \le \eps^a - \eps^{Br}_\parallel \ri \le \eps^{b} - \eps^{Br}_\parallel \ri \,,\\
\beta_{13} &=&  f_b + D_\parallel \le \eps^a - \eps^{Br}_\parallel
\ri \,, \\
\beta_{21} &=&
\nu_{21} \le \eps^a - \eps^{Br}_\perp\ri \le \eps^{b} - \eps^{Br}_\perp\ri \,,\\
\beta_{22} &=& \nu_{22} \le \eps^a - \eps^{Br}_\perp \ri \le \eps^b
- \eps^{Br}_\perp \ri + D_\perp \le 2
\eps^{Br}_\perp - \eps^a - \eps^b \ri -1 \,, \\
\beta_{23} &=&  f_b + D_\perp \le \eps^a - \eps^{Br}_\perp \ri
 \,.
\end{eqnarray}
Thus, the sought after   derivatives of  $\eps^{Br}_\perp$ and
$\eps^{Br}_\parallel$  finally emerge as
\begin{eqnarray}
 \frac{d \eps^{Br}_\parallel}{d
\omega} &=&  \frac{ \beta_{12} \beta_{23} - \beta_{22}
\beta_{13}}{\beta_{11} \beta_{22} - \beta_{12} \beta_{21}}
\,, \l{dexdw} \\
 \frac{d \eps^{Br}_\perp}{d
\omega} &=&  \frac{ \beta_{21} \beta_{13} - \beta_{11}
\beta_{23}}{\beta_{11} \beta_{22} - \beta_{12} \beta_{21}} \,.
\l{dedw}
\end{eqnarray}

\section{Numerical investigations} \l{Num_inv}

The consequences  of the theory presented in the previous section
are
 illustrated here by means of some numerical examples. We begin with
 the simplest case in \S\ref{section_sphere} wherein the pores  are
 spherical and the infiltrated porous material is accordingly
 considered to be an isotropic HCM. Then the effects of anisotropy
 are considered in \S\ref{section_spheroid} wherein the pores are
 taken to be spheroidal in shape. For the purposes of these numerical
 calculations, our attention is restricted to relative permittivity
 values which, at  optical frequencies,
  are attainable either using naturally-occurring materials or
 currently-available engineered materials. In \S\ref{closing} the
 results of implementing relative permittivity
 values which lie beyond the reach of present-day technology are
 commented upon.

\subsection{Spherical pores} \l{section_sphere}

If the pores  are spherical (i.e., $\rho = 1$) then the relative
permittivity dyadic characterizing the infiltrated porous material
reduces the the scalar form $\=\eps^{Br} = \eps^{Br} \=I$ with $
\eps^{Br} = \eps^{Br}_\parallel \equiv \eps^{Br}_\perp$. For $\eps^a
\in \lec 1.5, 5, 15 \ric$, the Bruggeman estimate $\eps^{Br}$ and
its derivative $d \eps^{Br} / d \eps^b$ are plotted versus $\eps^ b
\in \le 1, 3 \ri$ and $f_b \in \le 0, 1\ri$ in Fig.~\ref{fig1}. We
see that when $\eps^a = 1.5$, the Bruggeman estimate $\eps^{Br}$
varies approximately linearly with $\eps^b$ for all $f_b \in \le 0,
1\ri$. However,  the relationship between $\eps^{Br}$ and $\eps^b$
becomes increasingly nonlinear as $\eps^a$ increases. For $\eps^a =
1.5$, the derivative $d \eps^{Br} / d \eps^b$ increases in an
approximately linear fashion as the porosity $f_b$ increases,
regardless of the value of $\eps^b$. However, for $\eps^a = 5$  the
trend is rather different: here the values of $d \eps^{Br} / d
\eps^b$ peak at $f_b \approx 0.7$ and the height of this peak rises
as $\eps^b$ decreases. This peak in the value of $d \eps^{Br} / d
\eps^b$ becomes more pronounced as the value of $\eps^a$ increases.
 Indeed, at $\eps^a = 15$ this peak can be clearly observed even when
$\eps^b = 3$.

\subsection{Spheroidal pores} \l{section_spheroid}

\begin{figure}[!htb]
\begin{center}
\begin{tabular}{c}
\includegraphics[width=18.0cm]{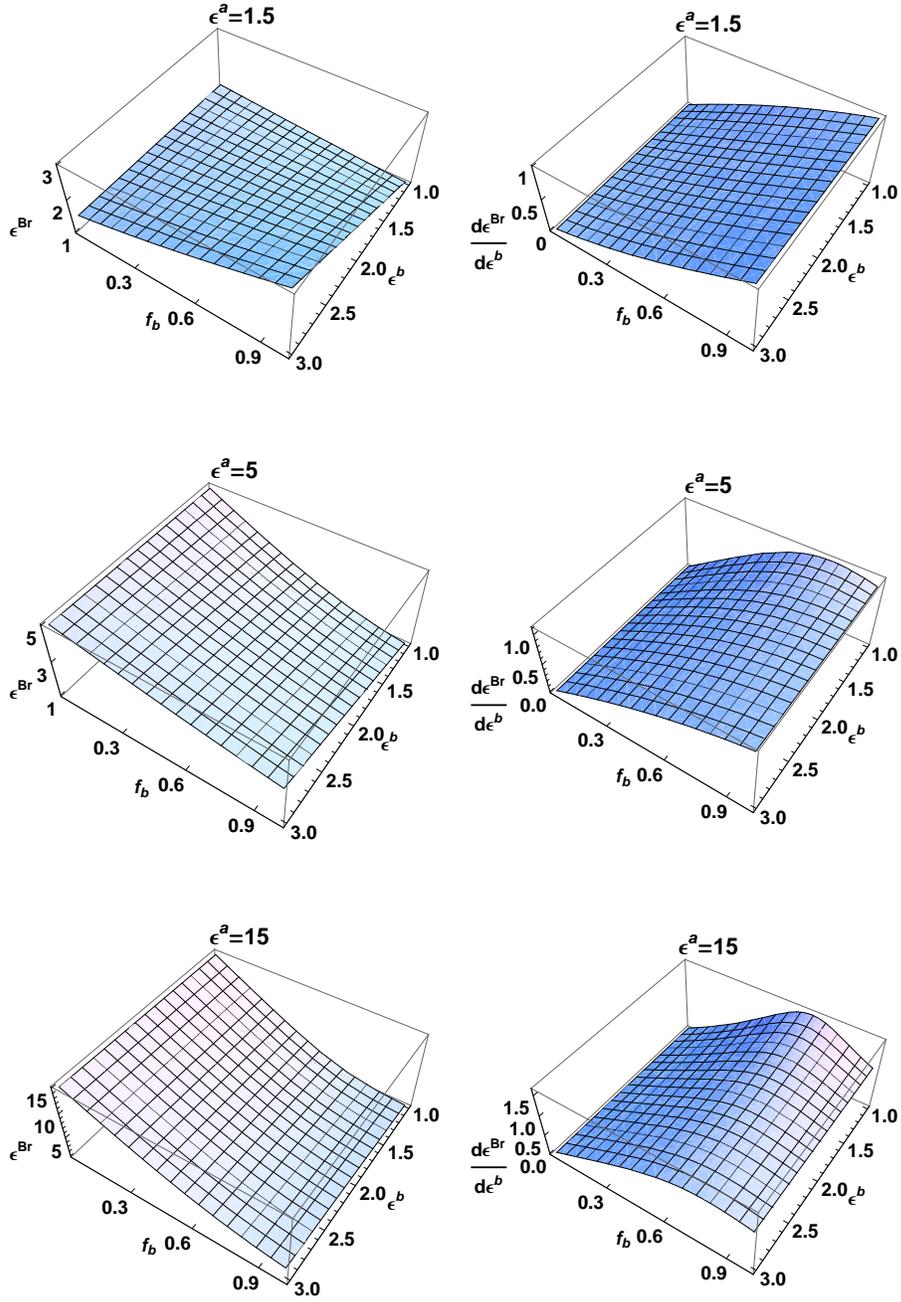}
\end{tabular}
\end{center}
 \caption{The  Bruggeman
estimate of the relative permittivity of the infiltrated porous
material $\eps^{Br}$  plotted versus the relative permittivity of
the infiltrating fluid $\eps^b \in \le 1, 3 \ri$ and the porosity
 $f_b \in \le 0, 1 \ri$,  for the relative permittivity of the
 porous
material $\eps^a \in \lec 1.5, 5 , 15 \ric$. Also plotted are the
corresponding derivatives $d \eps^{Br} / d \eps^b$.
 } \label{fig2}
\end{figure}

Let us now explore what happens when the pores are taken to be
spheroidal. Accordingly, the HCM representing the infiltrated porous
material is a uniaxial dielectric material. Following our findings
in \S\ref{section_sphere}, we fix $\eps^a = 15$ in order that the
effects of pore shape are more clearly appreciated. The Bruggeman
estimates $\eps^{Br}_{\perp,
\parallel}$ and their derivatives $d \eps^{Br}_{\perp,
\parallel} / d \eps^b$ are plotted versus
 $\eps^ b \in \le
1, 3 \ri$ and $f_b \in \le 0, 1\ri$ in Fig.~\ref{fig2} for $\rho =
10$. Both $\eps^{Br}_\perp$ and $\eps^{Br}_\parallel$ vary
relatively little as $\eps^b$ increases but both
decrease~---~$\eps^{Br}_\parallel$ approximately linearly and
$\eps^{Br}_\perp$  more nonlinearly~---~as $f_b$ increases. The
derivative $d \eps^{Br}_\parallel / d \eps^b$ increases
approximately uniformly as $f_b$ increases, for  $\eps^b \gtrsim
1.5$. However, the values of $d \eps^{Br}_\parallel / d \eps^b$ for
$\eps^b \lesssim 1.5$ are slightly peaked around $f_b \approx 0.9$.
 The plot of $d \eps^{Br}_\perp /
d \eps^b$ is  similarly peaked, but in this case the peak occurs at
$f_b \approx 0.5$, it is larger in height than the $d
\eps^{Br}_\parallel / d \eps^b$ peak, and it extends further into
the $\eps^b \gtrsim 1.5$ region. Also, the height of this $d
\eps^{Br}_\perp / d \eps^b$ peak  is substantially larger than the
corresponding peak in $d \eps^{Br} / d \eps^b$ observed in
Fig.~\ref{fig1} for $\eps^a = 15$.

\begin{figure}[!htb]
\begin{center}
\begin{tabular}{c}
\includegraphics[width=15.0cm]{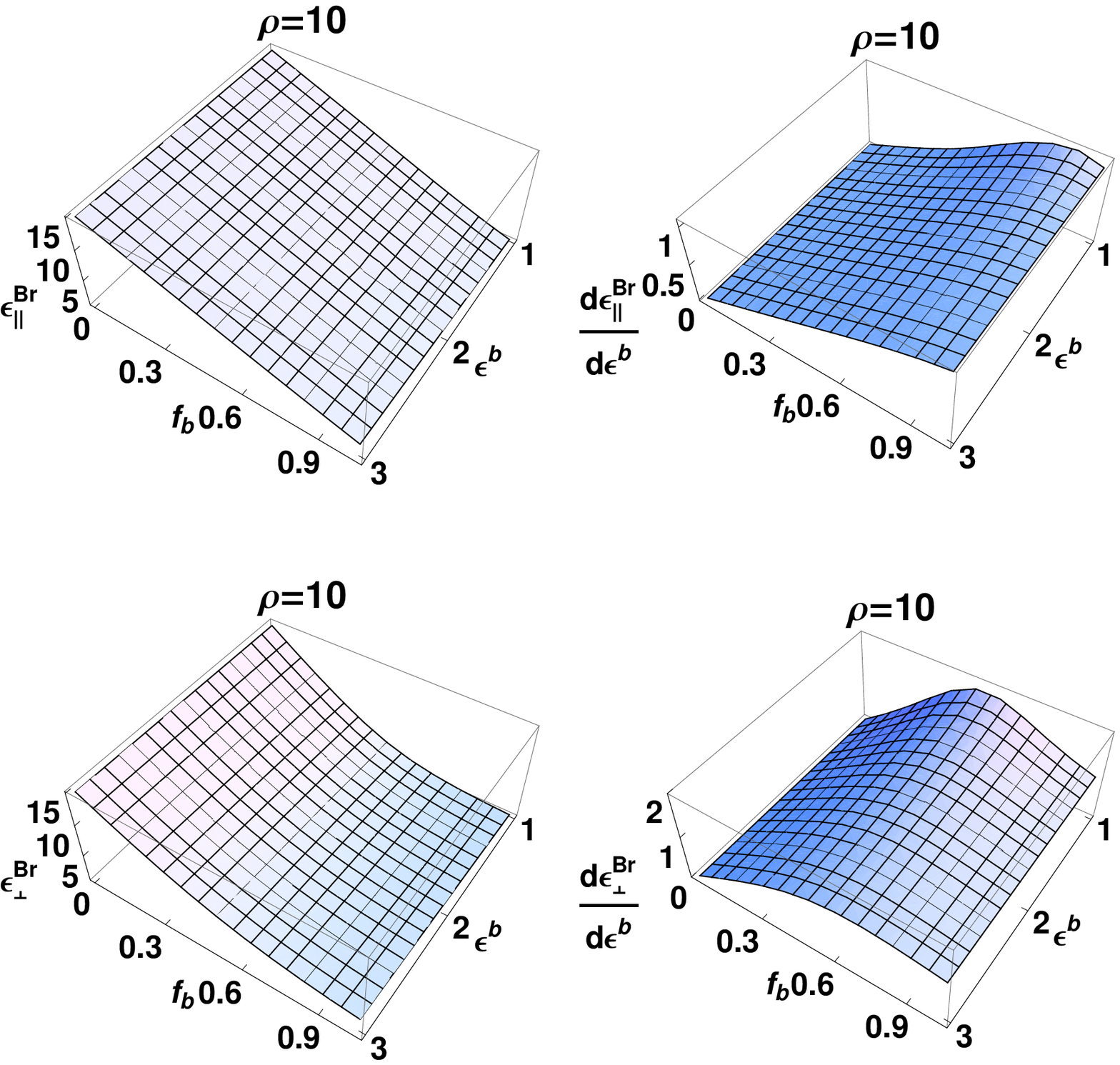}
\end{tabular}
\end{center}
\caption{The Bruggeman estimate of the relative permittivity
parameters of the infiltrated porous material $\eps^{Br}_{\perp,
\parallel}$  plotted versus the relative permittivity of the
infiltrating fluid $\eps^b \in \le 1, 3 \ri$ and the porosity $f_b
\in \le 0, 1 \ri$, for the pore shape parameter $\rho = 10$ and the
relative permittivity of the porous material $\eps^a = 15 $. Also
plotted are the corresponding derivatives $d \eps^{Br}_{\perp,
\parallel} / d \eps^b$.} \label{fig3}
\end{figure}

The pores   represented in Fig.~\ref{fig2} are prolate spheroids.
The corresponding case of oblate spheroids is represented in
Fig.~\ref{fig3}. The parameters for the plots  in Fig.~\ref{fig3}
are the same those in as Fig.~\ref{fig2} except that $\rho = 0.1$.
The plot of $\eps^{Br}_\parallel$ versus $\eps^b$ and $f_b$ in
Fig.~\ref{fig3} is very similar to the corresponding plot of
$\eps^{Br}_\perp$ in Fig.~\ref{fig2}; and likewise for the plots of
$\eps^{Br}_\perp$ in Fig.~\ref{fig3} and $\eps^{Br}_\parallel$ in
Fig.~\ref{fig2}. Also, the plots of the derivatives  $d
\eps^{Br}_\parallel / d \eps^b$ and  $d \eps^{Br}_\perp / d \eps^b$
versus $\eps^b$ and $f_b$ in Fig.~\ref{fig3} are similar to the
corresponding plots of
  $d
\eps^{Br}_\perp / d \eps^b$ and  $d \eps^{Br}_\parallel / d \eps^b$,
respectively, in Fig.~\ref{fig2}, albeit there are  qualitative
differences in the positions and shapes of the peaks in the
derivative plots.

\section{Discussion and closing remarks} \l{closing}

An analysis based on the Bruggeman homogenization formalism has
provided insights into the sensitivity of a generic porous optical
sensor. Specifically, for a porous  material of relative
permittivity $\eps^a$ infiltrated by a fluid of relative
permittivity $\eps^b$, we found that:
\begin{itemize}
\item  the
sensitivity is maximized when there is a large contrast between
$\eps^a$ and $\eps^b$;
\item if the contrast between
$\eps^a$ and $\eps^b$ is large, maximum sensitivity is achieved at
mid-range values of  porosity;
\item higher sensitivities may be achieved for $ \eps^b $ close to unity when $\eps^a \gg
1$, for example; and
\item higher sensitivities may be achieved by incorporating
elongated pores.
\end{itemize}

In \S\ref{Num_inv} the relative permittivities of the porous
material considered were $\eps^a \in \lec 1.5, 5, 15 \ric$. These
values correspond to  many common dielectric materials at optical
frequencies, with the largest value being close to the relative
permittivity of silicon, for example. The relative permittivities of
the infiltrating fluid were taken to be in the range $1 < \eps^b <
3$. This range is physically-realizable, with the largest values
corresponding to nanocomposite fluids developed for immersion
lithography \c{Langmuir} whereas values approaching unity may be
attained using water vapour \c{Steam}, for examples. We note that
ongoing rapid developments in engineered materials are bringing
relative permittivity parameter regimes, which were hitherto
unattainable, into reach \c{Simovski,Lederer}. For example,
relative permittivities in excess of 50 are now being reported for
engineered materials in the terahertz  frequency regime
\c{Shin_PRL,Nature_high_n}, while engineered materials with
positive-valued relative permittivities less than unity also appear
to be attainable \c{Alu,Lovat,Cia_PRB}. Accordingly, it is of
interest to consider how the sensitivities reported here would be
effected if rather more exotic parameter regimes were incorporated.
In further numerical studies (not presented in \S\ref{Num_inv}) it
was observed that increasing $\eps^a$ beyond 15 results in a steady
increase in sensitivity. Reducing the value of $\eps^b$ from unity
results  in a sharp increase in sensitivity. In this context let us
make a couple of parenthetical remarks: First, since the entire
analysis presented herein is isomorphic to the corresponding
scenario for magnetic materials (with relative permittivities
replaced by relative permeabilities throughout), we note that
relative permeabilities less than unity can be achieved by using
diamagnetic materials \c{Cook}.
 Second, the
 regime $\eps^ b< 0 $ with $\eps^a > 0$ (or vice versa) gives rise to  Bruggeman
  estimates of the HCM
 relative permittivity dyadic which are not physically plausible
 \c{Ag} and therefore this regime is avoided here.

The design parameters considered here were the relative
permittivities of the porous material and the infiltrating fluid,
along with the porosity and the shapes of the pores. In the
operation of such a generic optical sensor~---~at least in the most
straightforward mode of operation~---~it may be  envisaged that the
relative permittivity of the infiltrating fluid is a variable
quantity (which varies according to concentration of the agent to be
sensed) whereas the other design parameters remain fixed.
Accordingly, the value of the relative permittivity of the fluid
should be carefully chosen such that the sensitivity is maximized
over the expected range of concentrations  of the agent to be
sensed.

While our attention here has been confined to infiltrated porous
materials represented as uniaxial dielectric HCMs, a straightforward
extension of the presented analysis could accommodate biaxial
dielectric HCMs which represent certain STFs as optical sensors
\c{ML_IEEE_SJ,ML_IEEEPJ,ML_PNFA}.

Finally, the study described herein provides a step towards
 a comprehensive study of porous platforms for optical
sensing, which  incorporates such matters as  the
absorption/desorption phenomenons that dictate the response time and
the reversibility of the sensors.

\vspace{10mm}

%\noindent {\bf Acknowledgement:} The author thanks two anonymous
%referees for their helpful comments and suggestions.

\newpage

\newpage

\newpage

\newpage

\begin{figure}[!htb]
\begin{center}
\begin{tabular}{c}
\includegraphics[width=15.0cm]{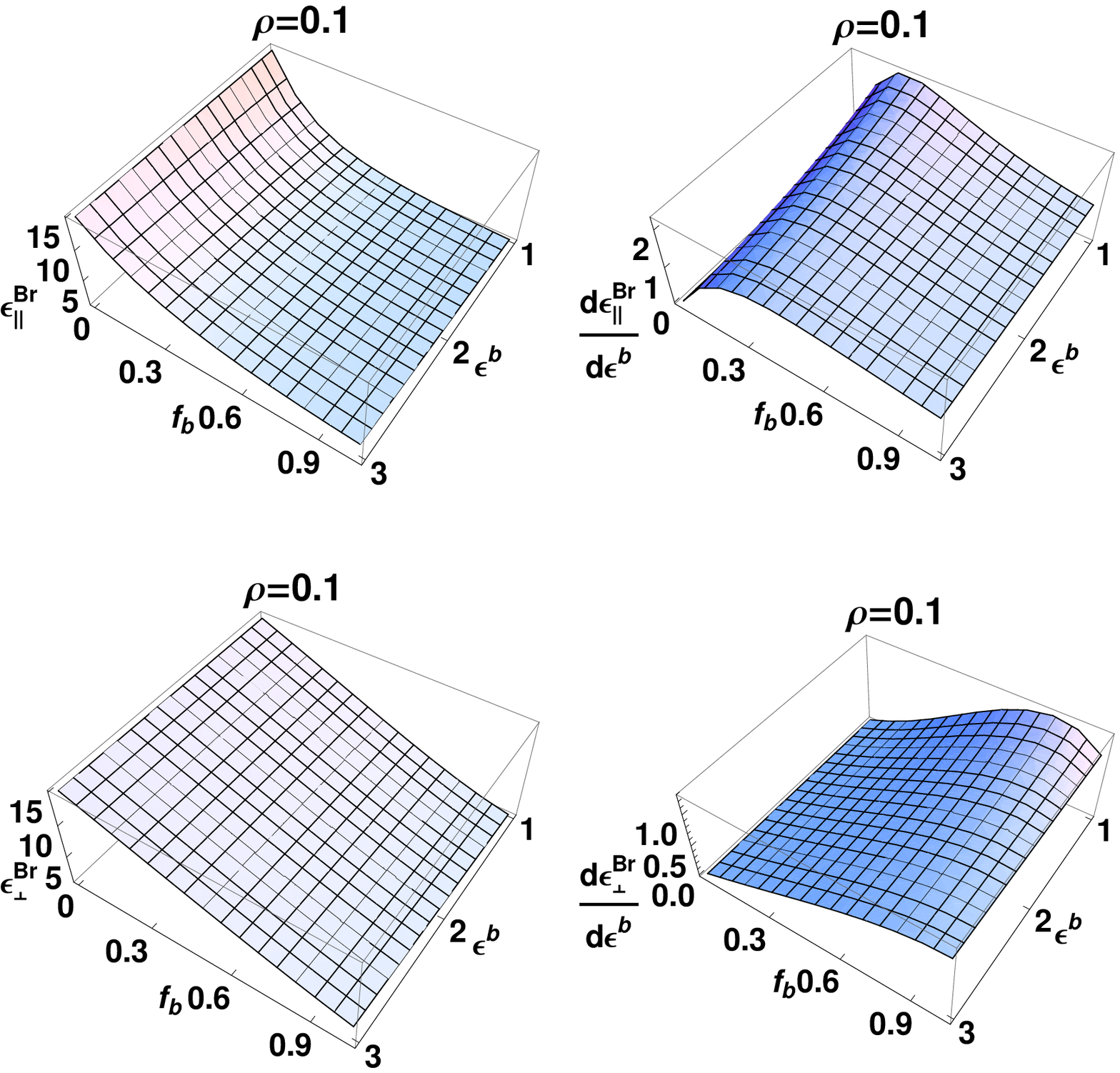}
\end{tabular}
\end{center}
 \caption{As Fig.~\ref{fig3} except that the pore shape parameter
$\rho = 0.1$. } \label{fig4}
\end{figure}


\begin{thebibliography}{99}

\bibitem{Stefano}
L. De Stefano, L. Rotiroti, E. De Tommasi, I. Rea, I. Rendina, M.
Canciello, G. Maglio, and R. Palumbo, ``Hybrid polymer-porous
silicon photonic crystals for optical sensing," \emph{J. Appl.
Phys.} {\bf 106}, 023109 (2009).

\bibitem{Pinet}
E. Pinet, S.  Dube, M.  Vachon-Savary, J.-S.  Cote, and M.
Poliquin, ``Sensitive chemical optic sensor using birefringent
porous glass for the detection of volatile organic compounds,"
\emph{IEEE Sensors J.} {\bf 6}, 854--860  (2006).

\bibitem{Sipe_OE}
J. J. Saarinen, S. M. Weiss, P. M. Fauchet, and J. E. Sipe,
``Optical sensor based on resonant porous silicon structures,"
\emph{Opt. Exp.} {\bf 13}, 3754--3764   (2007).

\bibitem{Homola2003}
J. Homola, ``Present and future of surface plasmon resonance
biosensors," \emph{Anal. Bioanal. Chem.} {\bf 377},  528--539
(2003).

\bibitem{AZL2}
I. Abdulhalim, M. Zourob, and A. Lakhtakia, ``Surface plasmon
resonance for biosensing: A mini-review," {\it Electromagnetics}
{\bf 28}, 214--242 (2008).

\bibitem{Scarano}
S. Scarano, M. Mascini, A. P. F. Turner, and M. Minunni, ``Surface
plasmon resonance imaging for affinity--based biosensors,"
\emph{Biosens. Bioelectron.} {\bf 25}, 957--966 (2010).


\bibitem{LMBR}
A. Lakhtakia, R. Messier, M. J. Brett, and K. Robbie, ``Sculptured
thin films (STFs) for optical, chemical and biological
applications," {\it Innovat. Mater. Res.} {\bf 1}, 165--176 (1996).

\bibitem{Lgas}
A. Lakhtakia, ``On determining gas concentrations using thin-film
helicoidal bianisotropic medium bilayers," \emph{Sens.  Actuat. B:
Chem.} {\bf 52}, 243--250 (1998).

\bibitem{L_Optik}
A. Lakhtakia, ``Enhancement of optical activity of chiral sculptured
thin films by suitable infiltration of void regions," \emph{ Optik}
{\bf 112}, 145--148 (2001).

\bibitem{STF_Book}
A. Lakhtakia and R. Messier, \emph{Sculptured Thin Films:
Nanoengineered Morphology and Optics}. Bellingham, WA, USA: SPIE
Press,  2005.

\bibitem{Messier}
R. Messier, V. C. Venugopal, and P. D. Sunal, ``Origin and evolution
of sculptured thin films,"  \emph{J. Vac. Sci. Technol. A} {\bf 18},
1538--1545 (2000).

\bibitem{Polo}
J. A. Polo Jr, ``Sculptured thin films," in \emph{Micromanufacturing
and Nanotechnology}, N.~P. Mahalik, Ed.. Heidelberg, Germany:
Springer, 2005,  pp.~357--381.

\bibitem{Horn}
S. M. Pursel and M. W. Horn, ``Prospects for nanowire
sculptured-thin-film devices," {\it J. Vac. Sci. Technol. B} {\bf
25}, 2611--2615 (2007).

\bibitem{Polo_PRSA}
J. A. Polo, Jr and A. Lakhtakia, ``On the surface plasmon polariton
wave at the planar interface of a metal and chiral sculptured thin
film,"  \emph{Proc. R. Soc. A} {\bf 465}, 87--107  (2009).

\bibitem{DPL}
Devender, D. P. Pulsifer, and A. Lakhtakia, ``Multiple surface
plasmon polariton waves," \emph{Electron. Lett.} {\bf 45},
1137--1138 (2009).

\bibitem{PML_JOSAB}
J. A. Polo, Jr, T. G. Mackay, and A. Lakhtakia, ``Mapping multiple
surface-plasmon-polariton-wave modes at the interface of a metal and
a chiral sculptured thin film," \emph{J. Opt. Soc. Am. B} {\bf 28},
2656--2666 (2011).

\bibitem{ML_IEEE_SJ}
 T. G. Mackay and   A. Lakhtakia, ``Modeling  chiral sculptured thin
films as platforms for surface-plasmonic-polaritonic optical
sensing,"  \emph{IEEE Sens. J.} {\bf 12}, 273--280 (2012).

\bibitem{Ward} L. Ward,  \emph{The Optical Constants
of Bulk Materials and Films, 2nd ed.}. Bristol, UK: Institute of
Physics,
 2000.

\bibitem{M_JNP}
 T. G. Mackay,
 ``Effective constitutive parameters of linear nanocomposites in the long-wavelength regime,"
  \emph{J. Nanophoton.} {\bf 5},  051001 (2011).

 \bibitem{ML_inverse_homog}
T. G. Mackay and A. Lakhtakia, ``Determination of constitutive and
morphological parameters of columnar thin films by inverse
homogenization,"  \emph{J. Nanophoton.} {\bf 4}, 041535 (2010).

\bibitem{ML_IEEEPJ} T. G. Mackay and  A. Lakhtakia,
 ``Empirical model of optical sensing via spectral shift
of circular Bragg phenomenon,"  \emph{IEEE Photonics Journal} {\bf
2}, 92--101 (2010).

\bibitem{ML_PNFA}
T. G. Mackay and  A. Lakhtakia,
 ``Modeling  columnar thin films as platforms for
surface--plasmonic--polaritonic optical sensing,"  \emph{Photon.
Nanostruct. Fundam.  Appl.} {\bf 8}, 140--149 (2010).

\bibitem{WLM}
W.~S. Weiglhofer, A. Lakhtakia, and B. Michel, ``Maxwell Garnett and
Bruggeman formalisms for a particulate composite with bianisotropic
host medium," \emph{Microw. Opt. Technol. Lett.} {\bf 15}, 263--266
(1997). Erratum: {\bf 22}, 221 (1999).

\bibitem{M97}
B. Michel,
 ``A Fourier space approach to the pointwise singularity of
an anisotropic dielectric medium," \emph{ Int. J. Appl. Electromagn.
Mech.} {\bf 8}, 219--227 (1997).

\bibitem{MW97}
B. Michel   and  W. S. Weiglhofer,
  ``Pointwise singularity of dyadic
Green function in a general bianisotropic medium," \emph{ Arch.
Elektron. \"Ubertrag.} {\bf 51}, 219--223 (1997). Erratum: {\bf 52},
 31 (1998).

\bibitem{MLD2005}
T. G. Mackay, A. Lakhtakia,  and R. A. Depine, ``Uniaxial dielectric
mediums with hyperbolic dispersion relations," \emph{ Microw.
  Opt. Technol. Lett.} {\bf 48}, 363--367 (2006).

\bibitem{Langmuir}
L. Bremer, R. Tuinier, and S. Jahromi, ``High refractive index
nanocomposite fluids for immersion lithography," \emph{Langmuir}
{\bf 25}, 2390--2401 (2009).

\bibitem{Steam}
P. Schiebener and J. Straub, ``Refractive index of water and steam
as a function of wavelength, temperature and density," \emph{J.
Phys. Chem. Ref. Data} {\bf 19}, 677--717 (1990).

\bibitem{Simovski}
C. R. Simovski, ``On electromagnetic characterization and
homogenization of nanostructured metamaterials," \emph{J. Optics
(UK)} {\bf 13}, 013001 (2011).

\bibitem{Lederer}
C. Menzel, T. Paul, C. Rockstuhl, T. Pertsch, S. Tretyakov, and F.
Lederer, ``Validity of effective material parameters for optical
fishnet metamaterials," \emph{Phys. Rev. B} {\bf 81}, 035320 (2010).


\bibitem{Shin_PRL} J. Shin, J. T. Shen, and S.  Fan, ``Three-dimensional metamaterials
with an ultrahigh effective refractive index over a broad
bandwidth," \emph{Phys. Rev. Lett.} {\bf 102}, 093903 (2009).

\bibitem{Nature_high_n}
M. Choi, S. H. Lee, Y.  Kim, S. B.  Kang, J. Shin, M. H. Kwak,
K.--Y. Kang, Y.--H. Lee, N. Park, and B. Min, ``A terahertz
metamaterial with unnaturally high refractive index," \emph{Nature}
{\bf 470}, 369--373 (2011).

%\bibitem{CPL}
%H. Xiao--Yang, C. Qi, L. Lin--Cui, Y. Chun, L. Biao, Z. Bang--Hua,
%and T. Chuan--Xiang, ``Nonresonant metamaterials with an ultra-high
%permittivity," \emph{Chin. Phys. Lett.} {\bf 28}, 057701 (2011).


\bibitem{Alu}
A. Al\`u, M. Silveirinha, A.  Salandrino, and N.  Engheta,
``Epsilon-near-zero metamaterials and electromagnetic sources:
Tailoring the radiation   phase pattern," \emph{ Phys. Rev. B} {\bf
75}, 155410 (2007).

\bibitem{Lovat}
G.  Lovat, P. Burghignoli, F.  Capolino, and D. R.   Jackson,
``Combinations of low/high permittivity and/or permeability
substrates for highly directive planar metamaterial antennas,"
\emph{IET Microw. Antennas Propagat.} {\bf 1},  177--183  (2007).

\bibitem{Cia_PRB}
M. N. Navarro-C\'ia, M.  Beruete, I.  Campillo, and M.  Sorolla,
``Enhanced lens by $\epsilon$ and $\mu$ near-zero metamaterial
boosted by extraordinary optical transmission," \emph{Phys. Rev. B}
{\bf 83}, 115112  (2011).



\bibitem{Cook}
J. J. H. Cook, K. L. Tsakmakidis, and O. Hess, ``Ultralow-loss
optical diamagnetism in silver nanoforests," \emph{J. Opt. A: Pure
Appl. Opt.} {\bf 11}, 114026  (2009).

\bibitem{Ag}
T. G. Mackay,
 ``On the effective permittivity of silver--insulator
nanocomposites,"
  \emph{J. Nanophoton.} {\bf 1},  019501 (2007).



\end{thebibliography}
\end{document}